\documentclass[twocolumn,showpacs,preprintnumbers,amsmath,amssymb]{revtex4}

\usepackage{subfigure}
\usepackage[dvips]{graphicx}
\usepackage{dcolumn}
\usepackage{bm}
 \newcommand{\eq}{\begin{equation}}
 \newcommand{\eqq}{\end{equation}}

\begin{document}

\title{Model for curvature-driven pearling instability in membranes}

\author{F. \surname{Campelo}}
\email{campelo@ecm.ub.es}
\author{A. \surname{Hern\'{a}ndez--Machado}}%
\affiliation{Departament d'Estructura i Constituents de la Mat\`{e}ria,\\
Facultat de F\'{\i}sica, Universitat de Barcelona\\
  Diagonal 647, E-08028 Barcelona, Spain}

\date{\today}

\begin{abstract}

A phase-field model for dealing with dynamic instabilities in membranes is presented. We use it to study curvature-driven pearling instability in vesicles induced by the anchorage of amphiphilic polymers on the membrane. Within this model, we obtain the morphological changes reported in recent experiments. The formation of a homogeneous pearled structure is achieved by consequent pearling of an initial cylindrical tube from the tip. For high enough concentration of anchors, we show theoretically that the homogeneous pearled shape is energetically less favorable than an inhomogeneous one, with a large sphere connected to an array of smaller spheres.

\end{abstract}

\maketitle


\textit{Introduction.--} Intracellular transport in eukaryotic cells is an essential process in cell biology \cite{alberts}. Deformations of different intracellular membranes, as those from the Golgi apparatus or the endoplasmic reticulum, lead to budding of transport vesicles, which may eventually undergo fission by the action of membrane proteins \cite{weigert99,kozlov06,voeltz07}. A membrane shape instability is, in some cases, the onset of those processes. Our work is thus motivated by these facts, aiming to give a simple physical interpretation of these morphodynamic effects in the cell interior. 


Dynamic instabilities in lipid vesicles and membranes have been widely studied experimentally. Tension-driven pearling induced by laser tweezers \cite{barziv94}, bilayer asymmetry \cite{chaieb98}, polymer anchorage \cite{tsafrir01}, or osmotic perturbations \cite{pullarkat06} has been reported, as well as budding and tubulation induced by polymer anchorage~\cite{tsafrir03}.

\begin{figure}[htbp!]
\centering
\vskip5mm
\includegraphics[angle=0,width=7.5cm]{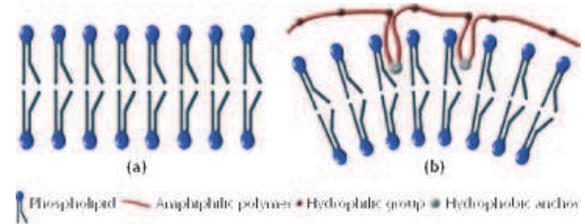}
\caption{Polymer wedge effect inducing a spontaneous curvature in a bilayer. A bilayer formed by one kind of lipids with zero spontaneous curvature tends to be flat (a). When a certain amount of anchor groups of an amphiphilic polymer gets stuck in the outer leaflet of the bilayer, a spontaneous curvature is induced (b).}
\label{fig:polymer_effect}
\end{figure}

A simple experimental model for membranes consists of a single component lipid bilayer and an amphiphilic polymer mimicking the shaping effect of anchored membrane proteins \cite{tsafrir01,tsafrir03}. When a certain amount of amphiphilic polymer is introduced in the vicinity outside a lipid vesicle, its morphology changes destabilizing the equilibrium vesicle shape \cite{tsafrir01,tsafrir03}. These experiments show that there is a coupling between polymer concentration on the membrane and local curvature. The polymer hydrophobic backbones anchor to the outer leaflet of the bilayer in order to minimize their hydrophobic interaction, acting thus as a wedge changing locally the curvature of the bilayer (see Fig.~\ref{fig:polymer_effect}).

Theoretical works on tubular vesicle morphologies have been reported in the literature. Some of them have restricted themselves to find stationary shapes \cite{miao91,bozic01}. A hydrodynamical explanation for the pearling instability occurring when a sudden tension is induced in the membrane by laser tweezers was given in Refs.~\cite{nelson95,goldstein96}. Some equilibrium models have dealt with shape instabilities due to the presence of anchoring molecules on the membrane \cite{tsafrir01,wang05}, but a dynamic model has not yet been proposed.


In this letter, we present a general method for dealing with morphological instabilities in membranes. In particular, we study the pearling instability induced by the anchorage of amphiphilic polymers on a lipid vesicle, in order to find and characterize the shape instabilities observed  experimentally \cite{tsafrir01}. We also show, motivated by our simulations, that the inhomogeneity in the pearl size which appears at high concentration of anchors on the membrane can be explained by energy considerations, consistently with the picture presented in \cite{tsafrir01}.


\textit{Theoretical treatment.--} A lipid vesicle subject to an effect which causes an asymmetry between the two leaflets of the bilayer vesicle can be elastically described by the Canham-Helfrich Hamiltonian \cite{canham70,helfrich73},
\eq\label{globalbending_sc}
\mathcal{H}_{\text{C-H}}=\frac{\kappa}{2}\int_{\Gamma}{\left(2 \mathsf{H}-\mathsf{C}_0\right)^2\text{d}S},
\eqq
where $\kappa$ is the bending modulus, $\mathsf{H}$ is the mean curvature, and $\Gamma$ is the vesicle surface. The asymmetry is introduced through the spontaneous curvature, $\mathsf{C}_0$. Minimization of this Hamiltonian under the constraints of constant area and volume yields the equilibrium shapes of vesicles~\cite{seifert97}.


To incorporate the effect of curvature generation by the anchored polymers, we assume a linear coupling between the spontaneous curvature and the polymer concentration \cite{leibler86}, $\mathsf{C}_0(\mathbf{x},t)=\mathsf{C}_0^{(0)}+\mathsf{C}_0^{(1)}\,\rho(\mathbf{x},t)$, where $\mathsf{C}_0^{(0)}$ is the bare spontaneous curvature, i.e. due to the asymmetry between the two leaflets of the bilayer before the polymer anchorage, $\mathsf{C}_0^{(1)}$ is the polymer-induced spontaneous curvature, and $\rho(\mathbf{x},t)$ is the local density of polymer. It has to be noted that here, the spontaneous curvature is, \emph{a priori}, a dynamic non-homogeneous function.


\textit{Phase-field model.--} The dynamics of a membrane considered as a surface is a paradigm of a moving boundary problem. A useful method for dealing with such problems is the phase-field approach (see \cite{gonzalezcinca04} for a review on surface tension-driven models). A phase-field model for taking into account the bending energy of lipid vesicles was presented in Refs.~\cite{campelo06,campelo_epjst_07}. The free energy functional proposed there was shown to be consistent with the already known stationary shapes of fluid vesicles.

We present here an extension of the phase-field model for the Canham-Helfrich bending energy derived in Ref.~\cite{campelo06}, in order to take into account a polymer-dependent spontaneous curvature. The free energy $\mathcal{F}[\phi,\rho]$, is a functional of the local polymer density, $\rho(\mathbf{x})$, and of a phase-field, $\phi(\mathbf{x},t)$, whose level-set $\{\phi(\bm{x})=0\}$ locates the interface.
\eq
\mathcal{F}[\phi,\rho]=\int_{\Omega}{\Phi^2[\phi ,\rho]\ \mathrm{d}\bm{x}},
\label{ansatz_sc}
\eqq
where
\eq
\Phi[\phi(\bm{x}),\rho(\bm{x})]=(\phi^2-1)\left(\phi-\epsilon\,C_0(\rho)\right)-\epsilon^2\bm{\nabla}^2\phi(\bm{x}),
\eqq
where $C_0$ is related to the spontaneous curvature as $\sqrt{2}\,C_0=\mathsf{C}_0$, and may depend in principle on any dynamic variable of the problem. Phase-field models are also called diffuse interface models, since the phase-field is a continuous albeit abruptly-changing function at the interface. The width of this diffuse interface is related to the small parameter $\epsilon$. In the sharp-interface limit, $\epsilon\to 0$, the phase-field model is equivalent to the equilibrium Canham-Helfrich Hamiltonian Eq.~\ref{globalbending_sc}. As in Ref.~\cite{campelo06}, we write an effective free energy functional $\mathcal{F}_{\mathrm{eff}}[\phi,\rho]=\mathcal{F}[\phi,\rho]+\int_{\Omega}{\sigma(\bm{x}) a[\phi]\mathrm{d}\bm{x}}$, where $a[\phi]=\frac{3}{2\sqrt{2}}\ \epsilon\left|\bm{\nabla} \phi\right|^2$, is the surface area density, and $\sigma(\mathbf{x})$ a local Lagrange multiplier which ensures local area conservation. The conservation of the inner volume of the vesicle is achieved via a relaxational conserved dynamics of the form $\frac{\partial \phi}{\partial t}=\bm{\nabla}^2\left(\frac{\delta\mathcal{F}_{\mathrm{eff}}}{\delta\phi}\right)$, leading to the dynamic equation for the phase-field,
\begin{eqnarray}\label{dyn.eqn}
\frac{\partial \phi}{\partial t}&=&2\bm{\nabla}^2\Big\{\left(3\phi^2-1+2\epsilon C_0(\rho)\, \phi\right)\Phi[\phi]-\epsilon^2\bm{\nabla}^2\Phi[\phi]\nonumber\\
&+&\epsilon^2\bar{\sigma}(\bm{x})\bm{\nabla}^2 \phi
\Big\},
\end{eqnarray}
where $\bar{\sigma}(\bm{x})=\frac{\sqrt{2}}{3 \epsilon}\sigma(\bm{x})$. A dynamic equation for this Lagrange multiplier is also needed. A first-order Lagrangian method \cite{bertsekas} is used to implement the conservation of the surface area.


A dynamic equation for the density field, $\rho(\mathbf{x},t)$, could also be straightforwardly introduced. However, for sake of simplicity, we are concerned here with the case of a homogeneous polymer distribution along the membrane, due to e.g. a global application of the polymer. A systematic study for non-homogeneous polymer concentration and polymer diffusion along the membrane si beyond the scope of this letter \cite{campelo_phd}.


Eq.~(\ref{dyn.eqn}) is a highly non-linear dynamic equation which must be solved numerically. We used a standard finite-difference scheme for the spatial discretization and an Euler method for the time-derivatives \cite{campelo06}. In all the results shown in this letter we used the value of the small parameter, $\epsilon$, equal to the mesh size of the lattice. The results are robust under variations of this parameter. Both the time and space discretizations are chosen in order to satisfy the Courant-Friedrichs-Lewy stability criterion \cite{bertsekas}.


\textit{Results and discussion.--} In the experiments by Tsafrir \textit{et al.} \cite{tsafrir01}, amphiphilic polymers are introduced in the bulk outside the vesicle, both globally and locally close to the tip of the tube. These molecules diffuse in the bulk until they come across the membrane, where they get stuck in such a way that their hydrophobic part anchors in the bilayer (in order to satisfy the hydrophobic interaction). Once a polymer is anchored in the bilayer, it diffuses along the membrane.

Here we consider the situation of global application of the polymer. In this case, we assume that the polymer concentration reaches a homogeneous stationary profile along the membrane almost immediately. The dynamic evolution is thus fully understood from the shape dynamics, so there is no need for a dynamic equation for the density field.

\begin{figure}[ht!]
\centering
\vskip6mm
\subfigure[Experimental Result (from Ref.~\cite{tsafrir01})]{\includegraphics[angle=0,width=6cm]{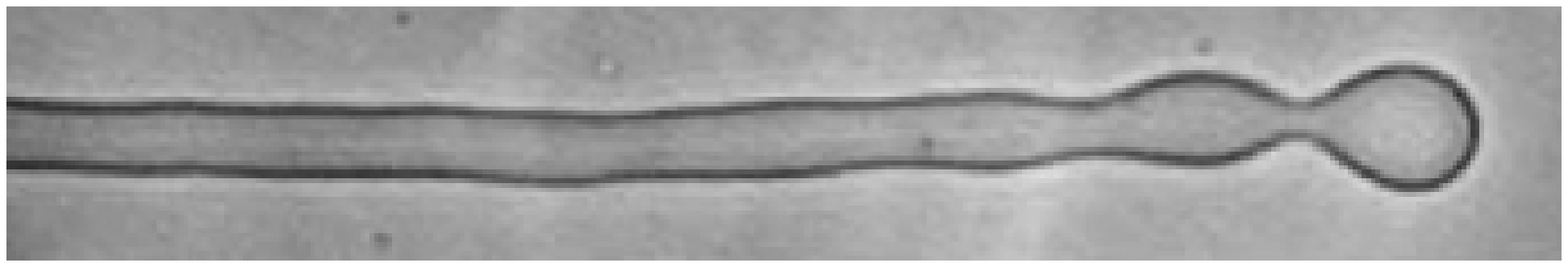}}
\subfigure[Phase--Field Simulation]{\includegraphics[angle=0,width=6cm]{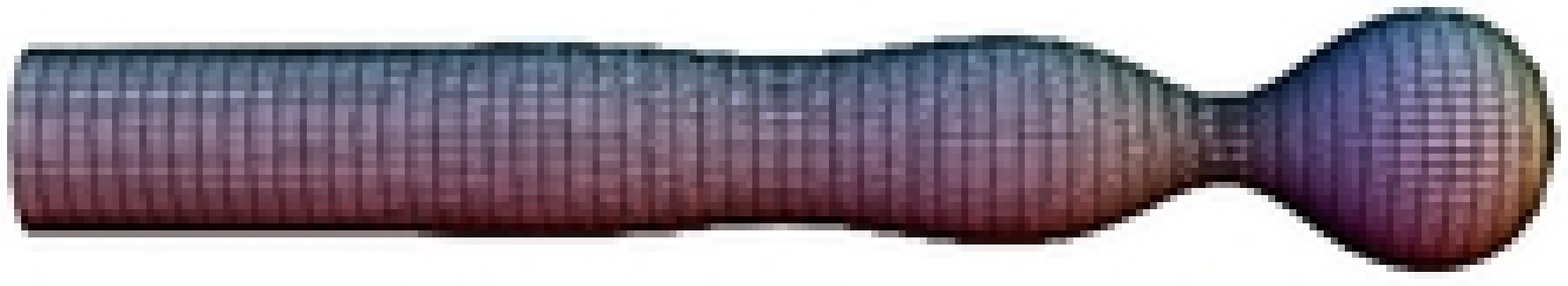}}
\caption{Onset of the pearling instability. Comparison of the experimental result from Ref.~\cite{tsafrir01} (a), and the phase-field simulation (b). It is important to note that in the simulation there is no fitting parameter, but we just let the system evolve from an initial tubular shape, under a relatively low homogeneous induced spontaneous curvature, $\mathsf{C}_0=0.48$, below the pearling instability limit.}
\label{fig:exp-phf}
\end{figure}

\textit{Onset of the instability.--} Tsafrir \textit{et al.} \cite{tsafrir01} studied the pearling instability in tubes whose length is much larger than their diameter. Within our model, we can find the shape of the tube at the onset of the instability (see Fig.~\ref{fig:exp-phf}). The experimental and predicted shapes are in good qualitative and quantitative agreement with each other. For instance, we can measure the ratio between the radius of the first pearl and the radius of the neck connecting it with the tube, and see that it gives a value of about $3$ in both the experiment and the simulation. We can thus assert that there is no need for an inhomogeneous polymer concentration on the membrane in order to start the instability, but it can just be triggered by the global change of the preferred curvature.

\begin{figure}
\centering
\vskip6mm
\subfigure[Experimental Result (from Ref.~\cite{tsafrir01})]{\includegraphics[angle=0,width=6cm]{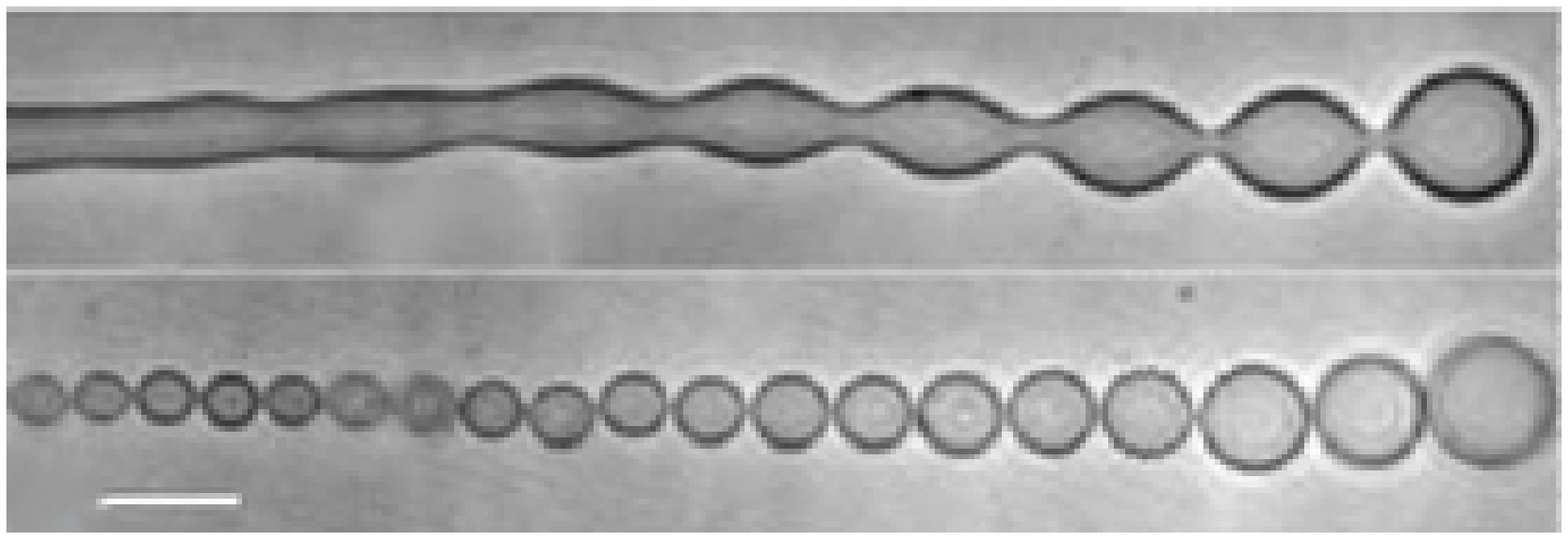}}
\subfigure[Phase--Field Simulation]{\includegraphics[width=6cm]{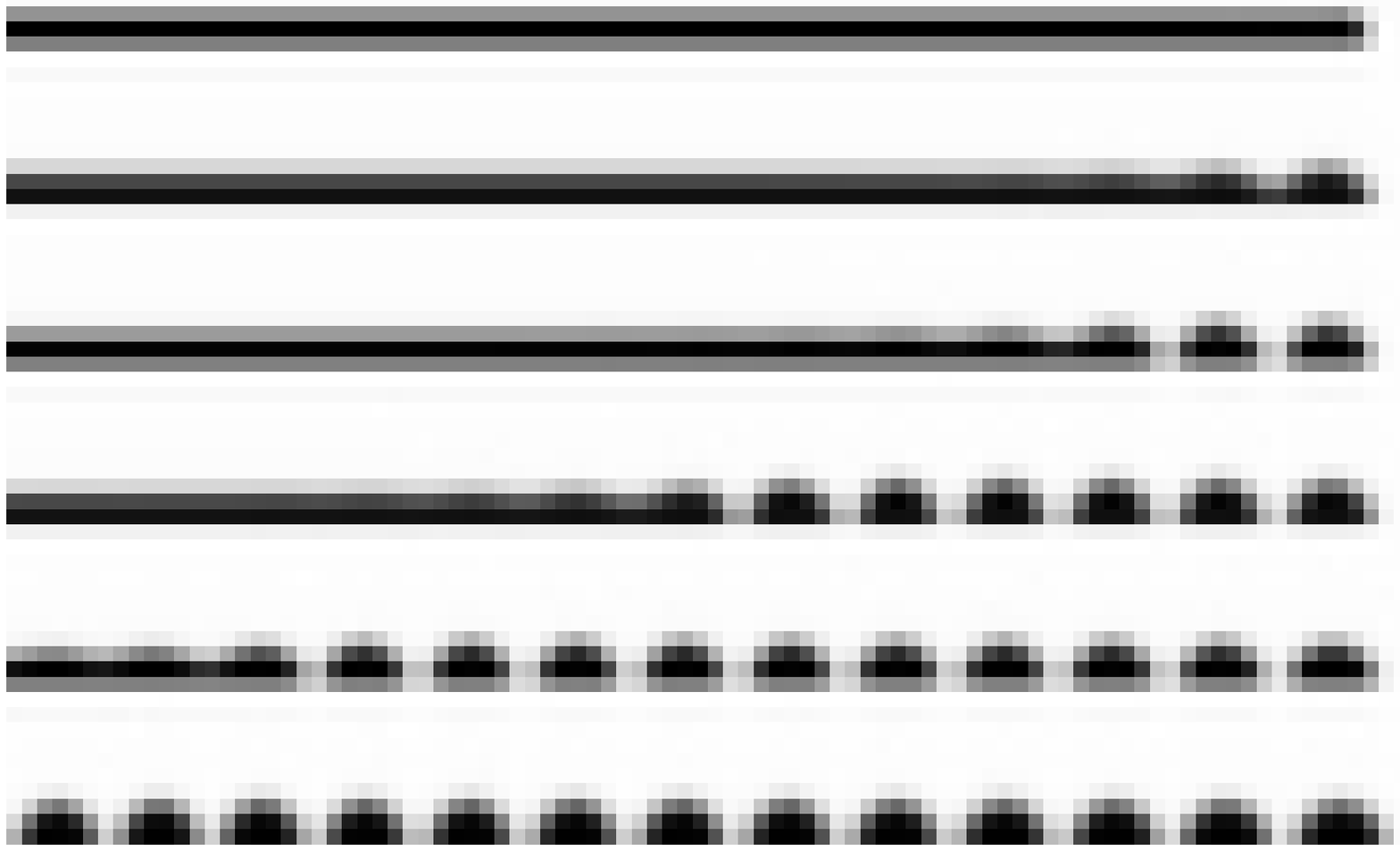}}
\caption{Dynamic evolution (time goes by from upper images to down) of a tube with an polymer concentration on the membrane such that $\mathsf{C}_0=0.68$. We show (a) the experimental results from \cite{tsafrir01} and (b) the phase-field simulation for comparison. Pearls start at points where the tube loses its perfect cylindrical geometry, namely the cap. The simulation has been performed on a $200\times 40$ axisymmetric lattice. No-flux boundary conditions at the lateral walls have been implemented.}
\label{fig:global_pearl_si}
\end{figure}

\textit{Low polymer concentration.--} Further addition of polymer solution increases the concentration of anchor chains in the membrane, therefore increasing the induced spontaneous curvature. Fig.~\ref{fig:global_pearl_si} shows the time evolution of a semi-infinite cylindrical tube with an endcap and the other end connected to a lipid reservoir. The homogeneous spontaneous curvature induced by the anchorage of the polymers made the initial cylindrical shape to be unstable and to create pearls.

We define the volume to area ratio, $\lambda=V/A$. For spontaneous curvatures between $\mathsf{C}_0=1/(2\, \lambda)$ and $\mathsf{C}_0=2/(3\, \lambda)$, Deuling and Helfrich \cite{deuling77} showed that there exist minimal surfaces, called Delaunay surfaces, which are global minima of the bending energy Eq.~(\ref{globalbending_sc}), for cylindrical shapes. The two limiting Delaunay shapes are a cylinder (for $\mathsf{C}_0=1/(2\, \lambda)$) and a set of spheres ($\mathsf{C}_0=2/(3\, \lambda))$. Unduloids, the one-parameter family of Delaunay shapes, interpolate smoothly between them. For higher spontaneous curvatures, the condition of vanishing bending energy cannot be fulfilled, and therefore the stationary shape corresponds to a non-vanishing minimum of the free energy. Our simulations show that, when the polymer solution is applied globally, i.e. when there is a simultaneous anchoring of polymer everywhere on the surface, the way of reaching a pearled structure from a cylindrical one is not by going through the family of equilibrium Delaunay shapes, but by a completely different dynamic evolution: creating pearls one by one from the tip of the tube (see Fig.~\ref{fig:global_pearl_si}(b)).

We checked how the free energy Eq.~(\ref{ansatz_sc}) varies in time seeing how each pearl formation is associated with an energy barrier which is crossed by thermal activation (due to numerical noise in our simulations). It is important to remark that there occurs no fission in the tube, but there is a narrow neck joining any pair of pearls, as seen in the experiments. The width of this neck is of the order of the mesh size. There is therefore no change in the topology of the tube. The tip of the semi-infinite tube is the point where pearls start when the polymer solution is applied both locally on the tip and globally \cite{tsafrir01}.

\begin{figure}
	\centering
	\includegraphics[width=8.5cm]{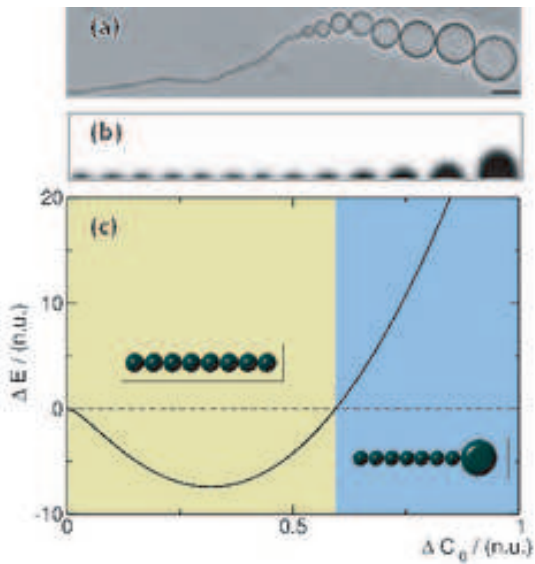}
	\caption{Inhomogeneous pearling. (a) Experimental result from Ref.~\cite{tsafrir01}, and (b) phase-field simulation using $\mathsf{C}_0=1.1$. (c) Plot of the energy difference $\Delta E$ between the bending energy corresponding to a set of spheres, and the one associated with a set of small spheres added to a bigger one, with respect to the increment of spontaneous curvature from the value $\mathsf{C}_0=2/3$, when equally-sized spheres have zero bending energy. For spontaneous curvatures bigger than a critical value, the homogeneous configuration is energetically less favorable than the inhomogeneous one. The area of the vesicle is finite, and chosen in such a way that the homogeneous pearled chain consists of 8 spheres. Magnitudes are measured in normalized units, given by $\kappa=1$ and $\lambda=1$.}
	\label{fig:endiff}
\end{figure}

\textit{High polymer concentration.--} It is seen in the experiments (see Fig.~\ref{fig:endiff}(a)) that when the concentration of anchored molecules on the membrane increases, there appears a gradient in the size of the pearls. The higher the concentration, the more pronounced the size gradient is. This is so because the higher the amount of anchored polymer on the membrane, the higher the induced spontaneous curvature. As we mentioned before, there are no energy-vanishing surfaces for $\textsf{C}_0>2/3$. Tsafrir \textit{et al.} \cite{tsafrir01} suggested that a shape consisting of a chain of equally sized Helfrich spheres connected to a larger sphere is, in terms of free energy, favorable to a chain of equally-sized spheres. Some pearls having the mean curvature equal to the spontaneous curvature are formed, but not all of them can fulfill this condition, under the volume and area constraints. Then, a larger sphere should form in order to keep these constraints. They argued that an inhomogeneity in the polymer concentration on the membrane (and thus in the distribution of the induced spontaneous curvature), may reduce the Helmholtz free energy $F=\mathcal{H}_{\text{C-H}}-T S$, $T$ being the temperature and $S$ the entropy.

However, we found that for higher values of the spontaneous curvature, a situation with a pearl size gradient may be energetically favorable to that of a chain of equally-sized pearls, even with the same spontaneous curvature all along the vesicle (Fig.~\ref{fig:endiff}).
It is interesting to note that for values of the spontaneous curvature just slightly higher than $2/3$, the homogeneous case keeps its stability against the inhomogeneous case. Once a critical value $\textsf{C}_0^{\phantom{0}\text{c}}$ is achieved, the homogeneous case destabilizes against the inhomogeneous one (see Fig.~\ref{fig:endiff}(c)). This critical value depends on the area of the vesicle or, in other words, on the length of the initial tube. This critical value decreases with increasing tube length, reaching $2/3$ in the case of infinite tubes. Therefore, we state that the inhomogeneous pearl size experimentally found by Tsafrir \textit{et al.} \cite{tsafrir01}, is not due to a inhomogeneous polymer distribution, but it is of a purely energetic nature.

\textit{Conclusions.--} In summary, a phase-field model for dealing with dynamic instabilities in vesicles has been studied in the case of curvature-driven pearling instability induced by the anchorage of amphiphilic polymers on the bilayer. We showed that the morphological changes reported in the experiments are explained by the generation of curvature by the anchors. We considered the situation in which polymer is applied globally, accounting for the main experimental results. Calculations in the framework of the Canham-Helfrich model showed that for a high enough homogeneous concentration of anchors, the homogeneous pearled shape is energetically less favorable than an inhomogeneous one, with a large sphere connected to an array of smaller spheres.


We are indebted to Joel Stavans for critical reading of the manuscript and for kindly providing us with his experimental results. We acknowledge financial support of the Direcci\'{o}n General de Investigaci\'{o}n under project No. FIS2006-12253-C06-05. F.C. also thanks Ministerio de Educaci\'{o}n y Ciencia (Spain) for financial support.

\bibliography{/home/felix/mypapers/bib}
\end{document}